\documentclass[12pt,a4paper]{article}
\usepackage{amsmath,amssymb}
\usepackage{exscale}
\usepackage{relsize}

\def\Example{\vskip3pt\noindent{\sc Example} }

\begin{document}
\parindent0.5cm

\title{On the relevance of the differential expressions 
 $f^2+f'^2$, $f+f''$ and 
$ff''-f'^2$
for the geometrical
and mechanical properties of curves }

\author{James Bell Cooper}
\date{}
\maketitle

\tableofcontents
\newpage
\begin{abstract}
The purpose of this article is to give a {\it pot-pourri}
of results on the mechanical and geometrical properties of curves
and explicit solutions to problems on  trajectories
of particles under suitable force laws.
The factor which unifies these rather disparate results is the
ubiquity of the expressions in the title.
We show how this explains in a unified fashion  a plethora of 
properties of a large class of special curves.
We also introduce two related ideas, the so-called $d$-transformation
key res
of a function and a new duality between trajectories for central
force laws and those for parallel laws.
The former will be our principal tool for obtaining
many results by a reduction to cases which can be solved by
elementary methods; the latter will allow us  to move back
and forth between results for central force laws and corresponding ones for
parallel laws.

Our treatment begins with an overview of the theory of 
the motion of a planet under a centripetal
law, in particular one which varies as a power of the distance
from the centre.  We derive
directly a criterion for 
a given orbit under a centripetal force to correspond to a power law,
This is essentially a modern version of Newton's method.  

It is implicit in Newton's treatment of the two-body problem that a single
orbit suffices to establish the force law and we give a short and elementary 
proof  that this is true in a strong form---namely that knowledge of 
the affine curvature
on an infinitesimal segment suffices.  

We then use the above criterion to give a 
detailed discussion of the Kasner-Arnold 
duality between power laws.  This involves the first two 
expressions of our title and  the $d$-transformation
mentioned above.  

We proceed to discuss two related themes in the differential 
geometry of special curves---the 
three 
differential expressions of the title and a  duality 
between curves which have natural representations 
as spirals
and those which can be conveniently described by parametrisations of the
form $(F(t),f(t))$, where $F$ is a primitive of $f$.   
We show that the former are 
ubiquitous in the computation of geometrical
or mechanical properties of such curves and display two families, the
MacLaurin spirals and catenaries, which are particularly 
rich in  these respects.

In the final section, we give explicit parametrisations of 
trajectories under a parallel force field,
emphasising again the case of power laws, and show that 
the differential expressions of the title and their stability properties
with respect to the $d$-transformation
provide a unified
approach.
We close with a  brief discussion of the theme of recreating curves from 
their curvature
functions.

\end{abstract}

\section{Introduction}

We start the technical part of this paper 
in section 2 below with an overview of one of the most fascinating topics
in the history of mathematical physics:  
the relationship between Kepler's and Newton's laws of 
motion.  
This was  the central
theme of  Newton's 
Principia.
We are particularly interested in the following two statements
about
an object (called the planet) which moves around a second one
(the sun) subject to a centripetal force (equivalently, in such
a manner that Kepler's second law holds---equal areas are swept out in 
equal times):

\noindent
I.  If the orbits are conic sections (more precisely ellipses, parabolas
or hyperbolas with the sun at a focus), then the force is inversely 
proportional to  ${r^2}$ where $r$ is the distance to the sun;

\noindent 
II.  The converse statement: if the planet moves under an inverse square
law, then the orbits are as above.

In the first edition of his monumental treatise [Ne]
Newton derived the statements of I (treating the three cases separately) and
stated that the converse , i.e., II, holds.  In later 
editions he added a proof of this converse, albeit one which is the 
subject of some controversy to this day.  

 Newton's solution to what is often referred to as the
Kepler problem is generally regarded as one of the key moments in the
history of science and represented the culmination of the work
of Tycho Brahe, who collected the data on the planetary motions,
and Kepler, who distilled his three laws from this data.

In addition, Newton developed a criterion which allowed him to deduce from
the geometrical form of an observed orbit whether the planet was moving under
a power law and to determine which power was involved.  He used this to 
investigate further exotic orbits which satisfy power laws not of the
Kepler variety.  Two of the more remarkable facts that he deduced were that
orbits consisting of circles which pass {\it through} the sun and conic sections
with the sun at the {\it centre} also arise from power laws (these cases
will be 
dealt with below).

It is implicit in Newton's work that the geometrical
form of a {\it single} orbit suffices to determine the force law.  
We shall demonstrate that if we observe an 
orbit of a planet moving under a central
field, then we can deduce the
force law from
its affine curvature. 

 Our statements above use  (as do all standard
treatments of the dynamics of planetary motion) several hidden
assumptions, since from
the observation of a single orbit we
can, in the most general situation, clearly only deduce information about
the force law at the points through which the planet passes.
Basically, we are assuming, in  addition to
Kepler's area law (equivalently, that the motion is determined
by a central force emanating from the sun):

\noindent
a) that the force is independent of time;

\noindent
b)  that the force depends only on the distance from its source (the sun)
, i.e., it is invariant under rotation around the sun.

\noindent
c) that the force depends on the radius in a (real) 
analytic fashion.  Hence, if we
observe a non-trivial orbit (that is, one which is not a circle with
centre at the sun) this will 
determine the nature of the analytic function on a non-degenerate 
interval and so in its entirety.  (In our statements below we shall
always assume that we do not have such 
a circular orbit in order to avoid this case---{\it any} 
force law which satisfies a), b) and c) clearly has circular
orbits as special solutions.)

Note that if we assume c) then it is not necessary to observe a complete
orbit to derive the force law.  Any non-trivial segment will suffice.

One of our central results can be expressed succintly as follows:

\begin{itemize}\item[A)] The force is proportional to a power of the
  distance from the central point (the \lq\lq sun'') if and only if
  the affine curvature $\kappa_{\text{aff}}$ is also proportional to a power of
  this distance. In addition, the  appropriate powers are related in a simple
  fashion, namely  $K\propto\frac 1{r^\alpha}\Leftrightarrow
  \kappa_{\text{aff}}\propto \frac 1{r^{\alpha +1}}$.

\item[B)] More generally, we have a force law $K\propto\phi(r)$ if and
  only if $\kappa_{\text{aff}}\propto \phi(r)/r$ where $\phi$ is a suitably 
smooth function
(we use the proportionality sign rather than equality to avoid
inserting constants). 
\end{itemize}

Curiously, we have been unable to find any relation between force laws
and affine curvature in the literature (with one exception---see
below). It is known that Newton used what in modern terms would be called 
curvature arguments but he can hardly have
used affine curvature since the concept was only
invented in the last century (see [Sc2] for a history of affine differential
geometry). 
There are various articles on Newton's use of curvature
(see, for example, [Br], [Co2]) but 
they clearly refer to the classical (Euclidean)
curvature which was available to him
as the inverse of the radius of the osculating circle.

There is, however,  one result on celestial mechanics which does
invoke (implicitly)  Euclidean curvature, namely the theorem of 
Hamilton [Ha]
which states that the motion is Keplerian , i.e., corresponds to an
inverse square law $K\propto\dfrac 1{r^2}$ if and only if the hodograph
is a circle , i.e., has constant curvature. (Recall that the hodograph
of the motion is the curve traced out by the velocity vector).

\noindent
We can incorporate this into our scheme as follows:
\begin{itemize}
\item[C)] The motion corresponds to a power law if and only if the
  Euclidean curvature $\kappa_{\text {hod}}$ of the hodograph is 
  proportional to a power of the distance from the corresponding
  planetary position to the
  origin. More precisely, we have $K\propto r^\alpha$ if and only if
  $\kappa_{\text {hod}}\propto r^{-\alpha-2}$. 
\end{itemize}

The relationship between the force law and the two curvatures can be
expressed succinctly in the formulae:

$$\kappa_{\text {hod}}.\kappa_{\text{aff}} \propto \frac 1 {r^3}, \qquad \kappa_{\text {hod}}.K
\propto \frac 1 {r^2}                                                                            ,\qquad \frac K{\kappa_{\text{aff}}} \propto r.$$

In the course of our investigations we stumbled in a natural manner
on the three-parameter family of functions
$$f(t) =p(\cos(d(t-t_0)))^{\frac 1d}$$
which  have the property that the corresponding orbits $rf(\theta)=1$
are induced by power laws.
Thecrucial fact is that the members of this family
have the special property that the result of an application
of any of the differential expressions of our title 
leads to a function which 
is proportional to a power of  $f$; this fact will be
the key to our treatment.  
More precisely, 
if $$f(t)=p(\cos d(t-t_0))^{\frac 1d}$$ then \\
$$f^2+f'^2=p^{2}f^{-2d+2},\quad f+f''=-p^{2}(d-1)f^{-2d+1},
\quad ff''-f'^2=-p^{2}df^{-2d+2}.$$

Thus the curves of the
form  $r^d \cos (d \theta) = 1$ all describe planetary orbits which
are induced by  power laws  (we omit the $p$ and $t_0$ since these
correspond  to the simple geometrical operations of dilation and
rotation respectively).  These curves were introduced by the 
renowned
mathematician Colin MacLaurin, who was aware of precisely this
fact. Further literature searches revealed 
that they had been investigated thoroughly in
the period of classical differential geometry (geometry of special 
curves).  Their equations are usually written in the form  
$r^n= \sin(n \theta)$ and they are known under the name of the MacLaurin  
spirals (sometimes sinusoidal spirals).
A central thesis of our paper is that these all arise from the simple case
$f(t)=p \cos(t-t_0)$ (a straight line) under what we call the 
$d$-transformation 
which associates to a function  $f$
a new one  $f_d$ where
$f_d(t)=(f(dt))^{\frac 1d}$. 
The latter's stability  properties
with respect to the differential expressions of our title 
explain in a unified manner many of these results.

The two standard treatises on special curves ([Go], [Lo]) devote extensive 
chapters to the development of some of the remarkable geometrical 
and mechanical properties of the MacLaurin spirals.

A further  theme of our treatment
is a duality between the curves discussed
so far and a second family of curves which have analogous properties
with respect to a force parallel to the $y$-axis, in  particular
one which is proportional to a power of the distance to the $x$-axis.

\noindent
 Three motivating examples are
\begin{enumerate}\item The cycloid $(t-\cos t,1+\sin t)$
    which is a brachistochrone and tautochrone for a constant force
    parallel to the $y$-axis, and a trajectory for a $\dfrac 1{y^2}$ law.
\item The circle $(-\cos t,\sin t)$, which is a geodetic for the
  Poincar\'e half-plane.
\item the catenary $y=\cosh x$, which is the form taken on by a hanging
chain.
\end{enumerate}

We remark that each of these curves can be parametrised in the 
form $(F(t),f(t))$ where $F$ is a primitive of $f$ which is in turn
a function of the above form , i.e., $p (\cos(d(t-t_0))^{1/d}$
(this is obvious in the first two examples,
the third one is more subtle).

The connection to the first class of curves is that there is a 
deep analogy between the properties of curves with
parametrisations $(F(t),f(t))$ (where 
$F$ is a primitive of $f$) and those of the spirals 
$r f(\theta) = 1$.
Both have remarkable
properties when  $f$  has the special form used in the definition
of the MacLaurin spirals.  It will be one of the main tasks of this
article to display the reasons for these two facts.

We shall also be interested in other families of curves associated
with a force law---brachistochrones, tautochrones or isochrones,
catenaries, elastica, geodetics for suitable metric tensors,
and for
light rays in  media where the index of refraction is proportional to a 
power of the
distance to a central point. For the sake of conciseness we
will simply refer to these curves as {\it trajectories}.

As a sample of these properties we mention that the MacLaurin spiral 
$r^d=\cos (d \theta)$:
\begin{enumerate}

\item
satisfies $\kappa\propto\dfrac 1{r^{d-1}}$ ($\kappa$ is the curvature);
\item
is an orbit for $K\propto\dfrac 1{r^{2d+3}}$;
\item
is a brachistochrone for $K\propto\dfrac 1{r^{2d+1}}$;
\item
is a catenary for $K\propto\dfrac 1{r^{d+2}}$.
\end{enumerate}
(see [Go]).

3) and 4) are examples of solutions of problems of the type: minimise
$\mathlarger\int r^\alpha\,ds$ for a suitable index $\alpha$.
This suggests that the
functions of the form 
$p(\cos (d (t-t_0))^{\frac 1d}$
will supply solutions to many concrete problems of
the calculus of variations, a fact which we shall verify and explain
from our unified point of view.

Our last theme is a description of the trajectories of particles moving
under parallel-force laws.
Recall that the  trajectories
of a particle moving under a given force law are the solution curves of
the differential equations
$$\frac {d^2x} {dt^2} = f(x(t),y(t)) \qquad\frac {d^2y}{dt^2}=g(x(t),y(t)),$$
where $(f(x,y),g(x,y))$ is the force at the point $(x,y)$.  The most interesting
cases are, of course, that of a central force, i.e., one of the form
$f(r)(\cos\theta,\sin\theta)$  (where $r$ and $\theta$ are polar
coordinates) which we discuss in section 2,
or  a parallel force of the form $(0,f(y))$.  Of particular 
interest are the cases where  $f$ is a power function, i.e.,
$K \propto r^\alpha$ in the central case or
 $K \propto y^\alpha$
in the parallel one.  Note that the family of trajectories is unchanged
if we multiply the force field by a {\it positive} constant.  Hence in
the above cases,
only the sign of the constant of proportionality affects the family
of trajectories, not its absolute value.

Of course, the famous result of Galilei (the trajectories for a constant
parallel force are parabolas) is, together with those of 
 Newton, one  of the key results in the 
history of physics.

One of Newton's
less well-known discoveries is the fact that a Dido circle (i.e., 
one which
is perpendicular to the $x$-axis) is a trajectory for a $y^{-3}$ law.  
This will follow from our results;
but we can go much further than Newton here.  Thus, we obtain {\it all}
trajectories for such a law and can show that these  curves are the only
{\it circles} which are trajectories for a $y^\alpha$ law.

The study of trajectories generated by force laws was an area of very
active research in the first half of the previous century and is associated
with Edward Kasner and his students. These investigators 
were interested in general
properties of families of curves which arise as such trajectories, rather
than in the explicit form of the families for concrete laws.
We think that it is of some interest to document the fact that we
can write down explicitly {\it all} the trajectories, using simple elementary
functions (and an integration), in the case of a parallel power law.

It is interesting that for the better-known case of a central power law, 
there are 
just three indices for which all of the trajectories can
be described explicitly using elementary functions: the Kepler case
$K \propto r^{-2}$, Hooke's law $K \propto r$ (where
the orbits are conic sections with centre at the origin) and the 
Cotes' spirals ($K \propto r^{-3}$) which we discuss in section 2.  
In general, the 
trajectories for a given force law form a three-parameter or $\infty^3$
family.  For the general central power law the MacLaurin
spirals mentioned above provide an explicit $\infty^2$ family of trajectories, 
but in the general case the remaining ones can as far as we know only be 
described
indirectly as far as we know 
(using functions which can be determined
implicitly after a quadrature).  In the Kepler case
for instance, MacLaurin's
family only picks up the parabolic orbits.

The special case of rectilinear motion turns out, perhaps surprisingly,
to be more intricate, and it is interesting to note that Newton, in his
Principia, devoted a whole section to this case, which he regarded as
a limiting case of the planar one (for a central force).  In this case, 
it is, of course, not the
geometrical form of the motion which is of interest but  its direct
description, i.e. formulae for the position as a function of time.
Here the results are less
satisfactory in the sense that we have to use not just elementary
functions but also the inverse of such a function.

In the final section we give descriptions of curves which satisfy the condition
that their curvature is proportional to a power of the distance from the 
$x$-axis.

Many of the results of this article are, of course, known; our contribution
has been to provide a unified approach.  However, we do give explicit formulae
and introduce special curves.

As mentioned in the text, we were led to consider the questions below
during the course of a cooperation with T. Russell and the late P.A. Samuelson
to whom we owe thanks for many fruitful discussions.  We would also
like to thank Iain Fraser who read through and commented on an earlier version
of the text (and, in particular, for his suggestingthe word \lq \lq quadrality'' as the appropriate substitute for \lq \lq duality'' for foursomes).

We now turn to the technical part of this article.  We begin with
a survy of the Kepler problem.
\section{The Kepler problem}

\subsection{Kepler's second law}

In this section we prove the basic fact that Kepler's second law (i.e.,  
that the area swept out by the planet in a given interval of time is constant)
is equivalent to the fact that the force is centripetal.
This well-known fact was proved by Newton in [Ne], but since it is central 
to our approach
we give a proof in the spirit of what follows.

We assume that the orbit of the planet has polar form  $r = f(\theta)$ for a 
smooth non-negative function  $f$.  Usually $f$ will be strictly positive, 
but in at least one case we will allow it to have a zero (i.e. for the orbit 
to pass through the origin).   Note that we are not assuming that $f$ is 
periodic, i.e. that the orbits are closed.  

In this situation,
the motion is determined when we know  $\theta$ as a function of time.  
The equations
of motion are then

$$x(t) = f(\theta(t)) \cos \theta(t), \quad y(t) = f(\theta(t)) 
\sin \theta(t).$$

\noindent The area swept out in the interval from $t_0$ to $t$ is

$$A(t) = 1/2 \mathlarger\int_{\theta(t_0)}^{\theta(t)} f^2(u)\, du .$$

\noindent The component of the acceleration perpendicular to the unit vector
$$(\cos\theta(t),\sin\theta(t))$$  is the scalar product of the
vector
$(-\sin\theta(t),\cos\theta(t))$ with the second derivative of
$(f(\theta(t)) (\cos \theta(t), \sin \theta(t))$ and an elementary
calculation shows that this is
$$2f'(\theta(t))\theta'^2(t) + f(\theta(t))\theta''(t).$$
On the other hand, we can use the fundamental theorem of calculus and 
the chain rule to  see that
$$A''(t)= f(\theta(t))f'(\theta(t)) \theta'(t)^2 + 
\frac 12 f^2(\theta(t))\theta''(t).$$

Hence  $A''(t)$ is a multiple of the component of the acceleration and so 
the vanishing of $A''(t)$  (which is just the analytical expression of 
Kepler's second law) is equivalent to the vanishing of the force component 
perpendicular to the vector from the sun to the planet.

\subsection{The inverse square law}
In order to prepare the reader for what follows, we consider briefly the Kepler
problem in its original form, i.e., for the inverse square law.  We shall
show shortly that if we write the equation of an orbit in the rather unusual
form  $r f(\theta) = 1$, then the force is proportional to a 
power of the distance if and only if $f$ satisfies a differential
equation of the form
$$f(\theta) + f''(\theta) = c f^\alpha(\theta)$$
for some constant $c$ and index $\alpha$. 
Then $K \propto r^{-2- \alpha}$.
(The left hand side of this equation is, of course, the first of our
differential expressions).
Thus an inverse square law corresponds to the case where $\alpha = 0$
and this gives a clue to why the Kepler universe is particularly stable: the
above equation is then linear, in fact it is simply
$f(\theta) + f''(\theta) = c$.  Of course this can be solved immediately,
and the reader will see that the result is precisely that of Newton.
Since the argument works in both directions, we have thus shown the
equivalence of I and II above, i.e., completed the Newtonian programme
with respect to the inverse square law.
Of course, we still have to derive the above equation and this we will now do.

\subsection{The Newton-Somerville equation}
Suppose that we observe one  planetary orbit, which
we now write as a polar equation
$r f(\theta) = 1$. The reason for using this form rather 
than the more usual one $r=f(\theta)$ employed above is that this leads to a 
significant increase in transparency and clarity in the computations.
In the Keplerian case, where the orbit is a conic section with the sun
at a focus,  the equation is---for a suitable choice of coordinate 
system---$r(1 + e \cos\theta) = 1$  where  $e$  is the eccentricity.  
In particular,
$0 < e < 1$  corresponds to  an elliptical orbit, $e=1$ to a parabolic, $e>1$
to a hyperbolic. The fact that $f(\theta)=1+e \cos \theta$
is a much simpler function than its reciprocal is a further clue as
to why our choice of equation is more natural in this context.

Consider now the configuration consisting of the dilations of the single
orbit and the rays going through the origin, i.e.,  the
level curves of the functions (in polar coordinates)
$u = r f(\theta), v = \theta$.

This is a so-called $S$-configuration (see [Co1])
and it is easy to calculate that the mapping $(x,y) \mapsto (U,V)$ where
$$U = \frac {r^2}2 f(\theta)^2 \mbox{ and } V = g(\theta)$$
is area-preserving and has our two systems as level curves,
where $g(\theta)$ is a primitive of $\dfrac 1{f(\theta)^2}$.

It follows  that if the area condition is satisfied, then (up to constants)
the time is given by
$t = g(\theta)$ and we can use this to get  $\theta$ as a function of $t$ 
by inverting $g$. In the interesting examples (e.g. in the Kepler case), 
one can compute  $g$  
explicitly,
but not its inverse.  Fortunately, as 
we shall see, for our purposes we shall not require this explicit
representation for $\theta$ as a function of time.

We have gone into this argument in some detail since it displays the
connection with the concept of an $S$-configuration. Of course, Kepler's
area condition leads directly to this expression for $t$.

We can now compute  a simple formula for $\dfrac{dt}{d \theta}$ and so 
(by the inverse 
function theorem) for
$\dfrac {d \theta}{dt}$ 
in terms of $f$ and its derivatives.  In fact,
as the reader will easily check,
$$\frac {d \theta(t)}{d t} = f^2(\theta(t)).$$

\noindent The motion is now (in Cartesian coordinates):
$$x(t) = \frac{\cos\theta(t)}{f(\theta(t))}, \quad  y(t) = 
\frac{\sin\theta(t)}{f(\theta(t))},$$
where $\theta(t) = g^{-1}(t)$. 

We now have the machinery we need to compute the derivatives of 
$(x(t),y(t))$.  
We  differentiate the vector function $(x(t),y(t))$  and use the 
chain rule to get the velocity 
vector as a function of time (in terms of $f$ and its derivatives).
The result is
$$v(t) = -(\sin \theta(t)f(\theta(t)) + \cos \theta(t) f'(\theta(t)),
- \cos \theta(t) f(\theta(t)) + \sin \theta(t) f'(\theta(t))).$$
\noindent Similarly, the acceleration vector is
$$a(t) = - (\cos \theta(t),\sin \theta(t)) f^2(\theta(t)) (f(\theta(t)) + 
f''(\theta(t))).
$$
We can immediately read off from the expression for the acceleration (which is,
of course, proportional to the force) that we have a power law if and only
if $f$ satisfies the above equation
$$f(\theta) + f''(\theta) = c f^\alpha(\theta).$$
(If we are only interested in the Kepler case, we can stop here since this 
closes the gap in our considerations above).
This equation is equivalent to one which can be found in [Ne]---see [Cha] for
a detailed discussion.  
Because of the central role that it will play in our considerations 
and of the relevance of the constants, we denote
it by $\text{ns}(c,\alpha)$ 
and call it the Newton-Somerville equation
(the first  statement in a modern
form of this criterion which we have been able to trace is in [So]).   
A recent reference
is [Po]---c.f. equation (6.4) there.

\subsection{A criterion for a power law}
If we differentiate this equation we get
$$f'(\theta) + f'''(\theta) = c \alpha f^{\alpha - 1}(\theta) f'(\theta)$$
and so we can eliminate $c$ by division to get

$$\frac {f'(\theta) + f'''(\theta)}{f(\theta) + f''(\theta)} = 
\alpha \frac {f'(\theta)}{f(\theta)}$$
, i.e.,
$$ \frac {f(\theta)(f'(\theta) + f'''(\theta))}{f'(\theta)
(f(\theta) + f''(\theta))} = \alpha. $$

We can reverse this reasoning to deduce that we have a power law if and 
only if
the derivative of the expression
$$\frac {f(f'''+f')}{f'(f + f'')}$$ vanishes, in which case $F \propto
 r^\beta$, where  $\beta = -2- \alpha$
and $\alpha$ is the (constant) value of $\dfrac {f(f'''+f')}{f'(f + f'')}$.

The equation for the existence of a power law is thus
$$f'(f + f'')(f f'' + f'^2 + f' f''') + f f'''' - (f f' + f f''')
(f'(f' + f''') - f''(f + f'')) = 0.$$
Summarising, the force satisfies a power law if and only if $f$
is a solution of this ODE and the power $\beta$ is then 
$-2 - \dfrac{f(f''' + f')}{f'(f+f'')}$. 

Due to the possibility of the functions in the denominator having zeroes,
one should perhaps regard this equation more as a heuristic principle.  Thus 
for a 
given $f$ one tests  (e.g. with Mathematica) whether it is a solution of the 
above equation.  If this is the case, one computes the appropriate quotient 
and verifies 
it for constancy. Possible zeroes of  $f'$ or $f + f''$ can then be 
investigated 
with {\it ad hoc} methods.

\subsection{Some applications}
Using this machinery we can instantly check whether a given orbit 
corresponds to a  power law.  We used it to check all of the
examples in Newton.  We then introduced parameters into the equations of 
such orbits
and experimented to find combinations which produce further examples.  
We mention two simple ones:

\Example
We used the above equations to determine when a circular orbit derives
from a power law  by testing the case of a circle with centre at 
$(a,0)$ and
radius  $1$.  We found that this satisfies a power law if and only if
$a=0$ or 
$a = \pm 1$.
Hence a circular orbit corresponds to a power law if and only if the sun lies
on the circle or at the centre. Both of these cases were discussed in 
Newton's Principia, but we have found no indication that he knew that these were the only ones.
This fact is of some historical interest since one of the hypotheses which
Kepler considered and rejected was that the orbit of Mars {\it was}
circular, but with the sun displaced from the centre.

Curiously, when we tried to extend this to the case of ellipses, i.e., to find
out if there are other possible positions of the sun (other than the known 
ones at the centre or at a focus) for a power law,
the computations turned out to be too
complicated to be completed by Mathematica.

Of the examples which we computed, we mention one which had interesting
consequences.
We tested the orbit
$$  f(\theta) = (a + b \cos (d \theta))^{1/d}$$
for a power law.  The above expression was cobbled together out of 
such examples as $f(\theta) = 1 + \cos \theta$,
$f(\theta) = (1 + \cos \theta)^{-1}$, $f(\theta) = \cos \theta$,
$f(\theta) = (\cos \theta)^{-1}$.  The reason for the unusual nature
of the dependency on $d$ will
become clear later.
Apart from known or trivial cases, this
provided us with two families of suitable functions, namely
$$f_d(\theta) = (\cos (d \theta))^{1/d}$$
and
$$g_d(\theta) = (1 + \cos (d \theta))^{1/d}$$

The family
$$f_d(\theta) = (\cos(d \theta))^{1/ d}$$  is well known
in the classical theory of curves and, as mentioned in the introduction,
will play a crucial role in our treatment.
 Its members satisfy the power law
$F \propto r^{-3 + 2 d}$.
The second family
$g_d(\theta) = (1 +\cos(d \theta))^{1/d}$
coincides essentially with the first one because of  the simple identity 
$(1+\cos d \theta)=2\cos^2(\frac{d \theta}2)$.

\subsection{Two simple cases}
We return to our central ODE
$$f + f'' = c f^\alpha.$$
with constants $\alpha$ and   $c$.    
In general, this equation is  non-linear.

However, there are two cases where it {\it is} linear, namely  $\alpha = 0$ 
(the Kepler case)
and  $\alpha =1$  (the $K \propto r^{-3}$ case), but 
it is linear for different reasons,this is evident from the 
different
role of the constant, and it is instructive to compare the solutions.

\noindent $\alpha=0$: the equation is $f + f'' = c$.  This is linear but 
inhomogeneous, and 
$c$ plays the role of the inhomogeneous term.  This equation can be solved
by elementary methods and the solutions are exactly of the form
$$f(\theta) = c + a \cos (\theta) + b \sin (\theta).$$
(This completes our analysis of the Kepler case since these correspond to 
the polar 
equations of conic sections with focus at the origin).
More precisely, if $c>0$
these
$f$  describe ellipses, parabolas 
or hyperbolas with the origin as a focus,
and the freedom in the choice of the constants  $a$  and  $b$  
(together with dilations) means that a curve specified
by a solution of the differential equation, i.e., a  curve of the form
$r f(\theta) = p$
where $f$ is a solution of the differential equation and  $p$  is a scaling 
factor, is a curve of the required type.  
For the remaining values of $c$ we get
uniform motion in a straight line ($c=0$), or hyperbolas 
(the Coulomb case $c<0$).

There is, however, one more case where this dependence can be computed, namely
the case  $\alpha = 1$.  Then the equation is
$f + f'' = c f$  or  $f'' = (1-c)f$.
Once again this is linear, but  $c$  plays an entirely different role, namely
in the coefficient of  $f$.  The solutions can be computed very
easily and are  logarithmic spirals, hyperbolic spirals       
and epispirals, depending on whether  $c < 1$, $c = 1$ or $c> 1$.

In this case we see that the class of orbits defined by the force law
depends in an essential manner on the constant of proportionality.  This case 
was 
settled by Cotes and the spirals which arise are known as Cotes' spirals.
For a modern treatment, see Whittaker [Wh].
  
As a final remark in this section we mention briefly the more general 
case of a 
force law which is a sum of two powers (also of considerable historical
interest).  These will appear again in the next section.

\Example:
Consider the equation
$$f + f'' = a f^\alpha + b f^\beta$$
which corresponds to a force law of the form 
$$F = \frac a {r^{2+ \alpha}} +   \frac b{ r^{2 +\beta}}.$$

\noindent The case $\alpha=0$ and $\beta = 1$ is interesting both for 
physical reasons
and also because it is the only genuine case of a sum of two powers 
which can be computed with ease.

\noindent In this case the differential equation is
$f + f'' = a + b f$ i.e. $f'' = (b-1) f + a$.

\noindent We distinguish the cases:

\noindent
a)  $b<1$.  The solution is

$$\frac a{1-b} + A e^{(1-b)^{1/2} \theta} + B e^{-(1-b)^{1/2} \theta};$$

\noindent
b)  $b>1$.  The solution is

$$\frac a{1-b} + A \cos[(b-1)^{1/2} \theta] + B \sin[-(b-1)^{1/2} \theta];$$

\noindent c) $b=1$.  The solution is

$$\frac {a \theta^2}2 + \theta + B.$$

\section{The Affine 
Curvature of an Orbit\\ Determines the Force Law} 

It follows from the formulae given above that the statement of the
title of this section holds.  Due to its intrinsic interest, we 
go into this in more detail.

\subsection{Further geometric quantities associated with orbits}
There are two further quantities which are determined by the geometry of the 
orbit and which turn out to be relevant---these are the curvature functions 
$\kappa$ and $\kappa_h$ of
the orbit and of the so-called hodograph, , i.e., the curve traced out by the
velocity vector.  The latter was introduced by Hamilton, who showed that
the presence
of an inverse square law is equivalent to the fact that the hodograph 
describes a circle, i.e., a curve with constant curvature.  These curvatures
can be
calculated from the above equations using the standard formulae from
differential geometry for the curvature of a parametrised curve.  Since
one has to differentiate the equations of motion three times to achieve
this task, one obtains potentially highly complicated expressions.  However,
a miracle takes place and they simplify to the tractable and significant 
formulae:
$$\kappa= \frac {f^3(f+f'')}{(f^2 + f'^2)^{3/2}},\quad 
\kappa_h = \frac 1{f + f''}.$$ 
(These formulae can be computed by hand---we also checked them 
using Mathematica).
The equation $f+f'' = \dfrac 1 {\kappa_h}$ can be regarded as a quantitative 
version of Hamilton's characterisation of the Kepler case, since the 
circular form 
of the hodograph (i.e. the constancy of $\kappa_h$) is equivalent 
to the validity of the equation $f+f'' = c$  for $f$ (we are tacitly assuming
that $f + f''$ is positive).

Thus the expression $f(\theta) + f''(\theta)$ which occurs in the central ODE
is the 
radius of curvature of the hodograph.
Since this expression  and the related one $f^2 + f'^2$ 
will be of crucial importance below, it is of interest that 
they can be 
expressed in terms of the curvatures $\kappa$ and $\kappa_h$, together 
with $f$, explicitly.  In fact,
$$f^2 + f'^2 = \frac {f^2}{(\kappa \kappa_h)^{2/3}}, \quad f+f'' = 
\frac 1{\kappa_h}.$$

Thus we see how two of the expressions from our title
arise in a natural way.
We now consider the affine curvature of the orbit.

\subsection{Affine curvature}  

Within the context of affine geometry, there are four concepts of curvature 
(affine curvature,
equi-affine curvature, central affine curvature and central
equi-affine curvature) depending on which type of geometry is
involved. These are characterised by the choice of Lie group to
define the geometry in question (in the spirit of Klein's Erlangen 
programme)---the affine group, the equi-affine group, the central
affine group or the central equi-affine group. For
completeness we recall briefly the definitions of these groups.

The affine group is the six-parameter group of all affine
transformations of the plane. (It will be convenient to use the 
classical terminology
and refer to this as an $\infty^6$-group).

The central affine group is the $\infty^4$ subgroup of those
affine transformations which leave a given point $S$ invariant ($S$ for
\lq\lq sun'')

The equi-affine group is the $\infty^5$ group consisting of those
affine transformations which are area-preserving. 

The definition of
the central equi-affine group  (an $\infty^3$-group) should now be 
self-explanatory. 

Each of these groups is associated with an appropriate notion of curvature.
In view of Kepler's second law it is natural to use the
last group in dealing with orbital mechanics. In order to avoid the unwieldy 
terminology \lq \lq central equi-affine
curvature'' we will refer to this simply as the affine curvature 
and denote it by $\kappa_{\text{aff}}$. 

We can give our main result a more intuitive content by recalling
that the affine curvature has the following direct geometric interpretation.  
We 
denote by $S$ the origin, by $P$ a typical point on the curve and
by $P'$ a neighbouring point.  We let the tangent to the curve at $P$
meet the ray $SP'$ at  $Q$.  Then the affine curvature at  $P$  is twice
the
limit, as  $P'$ tends to $P$, of the quotient of the area of the triangle
$P'PQ$ by the cube of that of $SPP'$ (see [Sc] for details).

\subsection{A computational proof}

We shall begin with a purely computational proof 
of the relationship between the force law and the affine curvature,
since this requires
no knowledge of affine geometry apart from the formula for $\kappa_{\text{aff}}$.
The formulae which we develop will also allow us to 
compute some simple illustrative examples. We will bring a
conceptual proof at the end of the article.

The required formula is 
$$\kappa_{\text{aff}}=\dot{\bf x}\wedge\ddot{\bf x}/(\bf x\wedge\dot x)^3$$
where ${\bf x}(t)$ is a parametrisation of the curve and we use the
Newtonian dots to indicate differentiation. The wedge product ${\bf
  x}\wedge{\bf y}$ of vectors ${\bf x}=(x_1,x_2)$, ${\bf y}=(y_1,y_2)$ is the
determinant $x_1y_2-x_2y_1$ of the corresponding $2\times 2$ matrix
(see [6] for the above formula).

If we plug the parametrisation
$(x(\theta),y(\theta))=\dfrac{(\cos(\theta),\sin(\theta))}{f(\theta)}$ into the 
formula for
$\kappa_{\text{aff}}$, a simple computation leads to the expression
$$\kappa_{\text{aff}}=(f(\theta)+f''(\theta))f^3(\theta).$$

The curvature of the hodograph is $\dfrac 1{f(\theta)+f''(\theta)},$
as can easily be computed by  using the
following parametrisation, which we include since it is of independent
interest (it allows us to compute the hodograph of any planetary orbit
from its geometric form):
$$(-f(\theta)\sin\theta-
f'(\theta)\cos\theta,f(\theta)\cos(\theta)-f'(\theta)\sin(\theta)).$$

\noindent
We remark that this  is the image of $(f'(\theta),f(\theta))$
under the reflection
matrix
$$-\left[\begin{array}{rr} \cos\theta&\sin\theta\\
\sin\theta&-\cos\theta\end{array}\right].$$

We remark that the MacLaurin spirals provide a large class of curves
whose affine curvature is proportional to a power of the distance from 
the origin and which are therefore orbits for power laws.

Remarkably, they also have the property that the {\it Euclidean} curvature
is  proportional to a power of the distance to the origin---a fact
that follows from the above relationship and the formula 
for the curvature of the orbit.

  Our results establish a relationship between the orbits of
power laws and classes of curves characterised by the fact that the
affine curvature is proportional to a a power of the distance to the
origin.  There are three cases where one can give simple explicit solutions
to the first problem and hence to the second one.  These are the Kepler
case of an inverse square law, the Hooke case ($K \propto r$) and
the case where $K \propto \dfrac 1 {r^3}$.  As noted above, the second
case is a well-known result, characterising the curves with constant
affine curvature as the conics with centres at the origin.  We have 
been unable to trace the other two in the literature and so state them
here for the record.  The conics with foci at the origin are characterised
by the condition $\kappa_{\text{aff}} \propto r^{-3}$.  The third family are 
the so-called
Cotes' spirals, which are characterised by the fact that 
their affine curvature
is proportional to $r^{-4}$.  They have the representation  
$r f(\theta)=1$ in polar coordinates,
where $f$ is a solution of an equation  
$f(\theta)+f''(\theta)= c f(\theta)$ for some $c$.  The form of the
solution depends on whether $c > 1$, $c=1$ or $c<1$.  Typical examples
are the spirals $r\cos (a \theta) =1$,  $r \theta = 1$, 
and logarithmic spirals.

\subsection{A conceptual proof}

We round off this section with a conceptual proof of the result on the affine 
curvature.  As is to be expected this
is very simple and follows more or less directly from the definition 
of the curvature notion that we use (which makes it all the more surprising
that this result has not been documented already).  Recall that  the 
classical ,i.e., Euclidean, curvature of a curve is defined as follows.
One introduces a special parametrisation, by arc-length,
and shows that the second derivative of the parametrised curve is then
proportional to the normal to the curve at the given point.  The curvature
is then defined to be the corresponding proportionality factor.  In the
case of the affine curvature one uses, as special parametrisation, the
area spanned by the ray from the origin to a given point $P$ on the curve.
In view of Kepler's second law, this parametrisation is proportional to 
time in the case of a planetary orbit. 
The second derivative of the parametrisation is then parallel to the ray
$OP$.  The affine curvature is defined to be the corresponding proportionality 
factor (once again, the precise argument and formulae can be found 
in [6]---see \S{}13).  Since the second derivative is proportional to the
acceleration, this explains the formulae used in our result.

\section{The Kasner-Arnold duality}

\subsection {Sinusoidal spirals and the $d$-transformation}
We now explain the
curious property of the sinusoidal spirals which we noted above, namely
that they {\it all} arise from power laws.
We begin with some remarks on these curves.
They are usually specified by their polar equations
$$r^n = \cos n \theta \qquad \text{or}\qquad r^n=\sin n \theta$$
where $n$ is a parameter which can range over the real numbers, the
second version explaining the nomenclature.  The first version will
be more convenient for our purposes (of course, the two are related
by a rotation).

This family was introduced by MacLaurin; particular choices of $n$ produce 
some familiar classical curves.  For example, 
$n=2$
is a hyperbola (with centre at the origin), $n=-1$ is a line, $n = -\dfrac 1 2
$ a parabola
(focus at the origin), $n = -\dfrac 13$ is Tschirnhausen's cubic 
(also called Catalan's trisectrix or L'Hospital's cubic), $n= \dfrac 1 3$ 
Cayley's 
sextic, $n= \dfrac 1 2$ is the cardioid and $n=2$ the leminiscate.

In a certain sense the family is generated by the two special cases $n=\pm 1$, 
i.e., the curves $r = \cos \theta$
and $r =\dfrac 1{ \cos \theta}$
(a unit circle tangential to the $y$-axis at the origin and the line
$x=1$ respectively). 
Since this fact is of some moment in what follows
we explain the process.  If we employ the substitutions 
$R = r^n$, $\Theta = n \theta$, then the equation $R = \cos \Theta$ reduces 
to that of the generic sinusoidal spiral.  Now the point with polar 
coordinates $(R,\Theta)$ is the 
image of $(r,\theta)$ under the mapping $z \mapsto z^n$ of the complex plane.  
In other words, the sinusoidal spirals are the pre-images of the circle  
$R = \cos \Theta$ or the line $R  \cos \Theta=1$ (depending on the sign of $n$)
under this mapping (alternatively the images under $z \mapsto z^{1/n}$).

At this point we remark that the correct setting for this discussion is not
the punctured plane (i.e., $\bf {R}^2$ or $\bf {C}$ without the origin)
but its universal covering surface, i.e., the Riemann surface of the
logarithmic function.  This avoids the usual difficulties with 
the non-uniqueness
or discontinuity
of the argument function.  However, in order to keep our dicussion elementary, 
we shall ignore this subtlety.

Since we have written our curves in the form  $r f(\theta) = 1$ (rather than 
  $r = g(\theta)$ which is more natural in the context of curve theory), 
we shall replace $n$
by $-d$ in the above equations, which now take the forms
$$r (\cos d \theta)^{1/d}=1$$
as before.

These considerations lead naturally to the concept of the 
{\it $d$-transformation}
$f_d$ of a function $f$, where
$$f_d(t)=(f(dt))^{\frac 1d} .$$
As we shall see shortly, this transformation has particular
stability properties with regard to the differential expressions of our 
title and
this will explain many phenomena on the mechanical and geoemetrical 
properties of special curves.

We now return to our central equation
$$f(\theta) + f''(\theta)= c f(\theta)^\alpha$$,
which corresponds to a force law with $\beta = -2 - \alpha$.  
Owing to its significance and also the 
importance of the constants $\alpha$ and $c$, we denote this equation by 
$\mbox{ns}(c,\alpha)$.

\subsection{The general situation}
The starting point of the ensuing discussion was a fact which struck us 
on reading Newton's Principia, in the version of Chandrasekhar [Ch].  One of
the fascinating results he obtains is that when the orbits are conic sections,
but with the Sun at the {\it centre}, then this also satisfies a power
law, albeit with $\beta = 1$.  These orbits are the images of the Kepler
ones under the mapping $z \mapsto z^2$ (cf. Arnold [Ar]).
This so-called duality, which apparently was first studied systematically
by Kasner although we have been unable to find a precise reference,
has been investigated in detail by many authors, see, 
for example,  the same reference.
The two families above suggest that there is a certain stability for power 
laws under
transformations of the form $z \mapsto z^n$  and it is this phenomenon which
we now investigate.  Simple computations  show that this is not
the case in general, but that for certain special situations, including
the ones mentioned above, unexpected coincidences occur in the formulas
and this explains the above facts.

Thus our central question is: suppose that $f$ satisfies 
$\mbox{ns}(c,\alpha)$.  Does $g$ also satisfy such an equation 
(with different constants) where $g(\theta) = (f(d \theta))^{1/d}$, 
i.e., $g$ is the $d$ transform of  $f$?
Our above remarks indicate that there is a result of this sort for the
special cases:

\noindent a) $f(\theta) = \cos \theta$ ($d$ arbitrary);

\noindent b) $f(\theta) = 1 + e \cos \theta$ ($d$ = 2).

As mentioned above, the expressions  $f(\theta) + f''(\theta)$
and $f(\theta)^2 + f'(\theta)^2$ will
play an important role in our computations.

\noindent 
In the above special cases their values are  are:

\noindent
a) $f(\theta) + f''(\theta)=0, \quad f(\theta)^2 + f'(\theta)^2=1$;

\noindent
b) $f(\theta) + f''(\theta)=1, \quad f(\theta)^2 + f'(\theta)^2=2 
f(\theta) + (e^2 - 1)$.

\noindent
Now there is a relationship between these two expressions.  Indeed

$$ \frac d {d \theta} (f(\theta)^2 + f'(\theta)^2) = 2 f'(\theta)
(f(\theta) + f''(\theta)).$$
(This will be discussed in  more detail below.)

It follows that if $f$ is a solution to $\mbox{ns}(c,\alpha)$,
then  
$$ f(\theta)^2 + f'(\theta)^2 = 
\frac {2 c f(\theta)^{\alpha + 1}}{\alpha + 1} + b$$
for some constant $b$  (or 
$ f(\theta)^2 + f'(\theta)^2 = 2 c \ln f(\theta) + b$ if $\alpha = -1$).
We emphasise that $b$ depends on the particular orbit that we are 
examining, within a universe
with power law $K\propto \dfrac 1{r^{2 + \alpha}}$ and \lq \lq gravitational 
constant'' $c$.    Those orbits with $b=0$ will 
be of particular interest below.  For the case of a $\dfrac 1 {r^2}$ law, 
these are precisely the parabolic orbits.

A simple computation shows that if
$g(\theta) = f(d \theta)^{1/d}$ (and so $f(d \theta) = 
g(\theta)^d$), then, whenever $f$ is a solution of $\mbox{ns}(c,\alpha)$,
we have
$$g^2(\theta) + g'^2(\theta) = \dfrac {2 c}{\alpha + 1}
g^{2 + \alpha d - d}(\theta) + b g(\theta)^{2-2d}$$
for the constant $b$ above
and so 
$$g(\theta) + g''(\theta) = \frac {c(2+ \alpha d - d)}{\alpha + 1}
g(\theta)^{1 + \alpha d - d}
 + b(1-d)  g(\theta)^{1-2 d}$$
for $\alpha \neq -1$.  (The case $\alpha = -1$ must be dealt with 
separately).

From these equations we can garner a wealth of information. 
Firstly we see that   the
dual of a power law is always the sum of two powers 
(again for $\alpha \neq -1$).  

A particularly interesting case is when $\alpha = 1$ and $n = \dfrac 12$.
Then we see that a $\dfrac 1r$ law is dual to one of the form
$$\frac {c_1}{r^2} + \frac {c_2}{r^3}$$
as discussed above.

There are three special situations where the dual collapses to a power law.

\noindent a) $c=0$, i.e., the case of straight line motion.  Then 
$g$ is a solution of $\mbox{ns}((1-n)b,1-2n)$ for any $n$.  This explains 
the case a) above where $f(\theta) = \cos \theta$ and  the dual curves 
are the sinusoidal spirals.

\noindent b) $b=0$.  Then $g$ is a solution of
$$ \mbox{ns}\left(\frac {c(2 + \alpha n - n)}{\alpha + 1},1 + \alpha n - n
\right) $$
for any $n$.

This means that $z \mapsto z^d$ transforms a  $\dfrac 2d - 3$ power 
law into a $2d - 3$ law.
This can be used to investigate the duality phenomenon in more detail.  Since 
a great deal has been published on this topic we will not dwell on it  but 
remark that the case where $d = \dfrac 12$ corresponds to
Newton's result  on the duality between conic sections
with sun at the centre and such sections with the Sun at a focus, i.e., with
$\beta = 1$ and $\beta = -2$ respectively.
Another interesting case is where $n=-1$.  This maps the case $\beta = -5$
onto itself and thus shows that this case is self-dual (cf., Arnold [Ar]).

\noindent

\noindent
As a final remark we note that the computations for the
case  $\alpha = -1$  are slightly different owing to the presence of the 
logarithmic term and we leave them to the reader.  As a consequence we see that
a dual in this case is never a power law (except in the trivial case $d=1$).

This is significant as it allows us to eliminate the third possibilty for 
dualities.  

\noindent c)  where the two powers $1 +\alpha d - d$ and 
$1 - 2 d$ in the above equation coincide.  However, this is the case where 
$\alpha = -1$ and so no non-trivial duality arises in this manner.

\section{A Duality 
between Trajectories for Central and Parallel Force Laws}

As we have seen, the members of
the three-parameter family 
$$f(t) = p (\cos (d(t-t_0))^{1/d}$$ 
are solutions of $\text{ns}(c,\alpha)$ 
for various values of the parameters $c$ and $\alpha$, and so 
the curves of the
form  $r^d \cos (d \theta) = 1$ all describe planetary orbits which
are induced by  power laws  (we omit the $p$ and $t_0$ since these
correspond  to the simple geometrical operations of dilation and
rotation respectively).

The main theme of this section is a duality between the curves discussed
so far and a second family of curves which have analogous properties
with respect to a force parallel to the $y$-axis, in  particular
one which is proportional to a power of the distance to the $x$-axis.

\noindent
We recall the three motivating examples 
\begin{enumerate}\item The cycloid $(t-\cos t,1+\sin t)$,
    which is a brachistochrone and tautochrone for a constant force
    parallel to the $y$-axis, or a trajectory for a $\dfrac 1{y^2}$ law.
\item The circle $(-\cos t,\sin t)$, which is a geodetic for the
  Poincar\'e half-plane.
\item The catenary $y=\cosh x$, which is the form taken on by a hanging
chain.
\end{enumerate}

(We remark that each of these curves is a solution of a classical problem
of the Calculus of Variations, for which see below. 2. and 3. are also
trajectories for $\dfrac 1{y^\alpha}$ laws).
The connection to the first class of curves is that there is a 
deep analogy between the properties of curves with
parametrisations of the form $(F(t),f(t))$ (where 
$F$ is a primitive of $f$) and those of the spirals 
$r f(\theta) = 1$.
Further, both have remarkable
properties when  $f$  has the special form used in the definition
of the MacLaurin spirals.  It will be one of the main tasks of this
article to display the reasons for these two facts.

We can summarise the above remarks as follows:

\begin{itemize}\item[A)] The curves of the first  family are \lq\lq
  spirals'' with equations of the form $rf(\theta)=1$ for suitable
  functions $f$ of one variable.
\item[B)] The curves of the second family have parametrisations of the
  form $(F(t),f(t))$ for a suitable function $f$ with $F$ a primitive of $f$.
\item[C)]
Further, when the functions $f$ which
 occur have the form $f(\theta)=(\cos d\theta)^{\frac 1d}$ for some
 parameter $d$ (or can be written in this form by 
 simple transformations), then the curves have remarkable properties.
As we shall see shortly, this is because these functions are the
solutions of three particular ordinary 
differential equations involving the differential expressions of our title,
a fact  of some
consequence for the mechanical properties of the corresponding curves. 
\end{itemize}

\subsection{Curves of the form $(F(t),f(t))$}
We therefore consider in more detail parametrised curves of the
form $$(x(t),y(t))=(F(t),f(t))$$
where $F$ is a primitive of $f$.

In a certain sense, every generic plane curve with
parametrisation $(x(s),y(s))$ can be reparametrised in the form 
$\left(\mathlarger\int^tf,f\right)$. We simply set 
$t= \mathlarger\int^s\dfrac{x'(u)}{y(u)}\, du$  which means
that
$\dfrac{dt}{ds}=\dfrac{x'(s)}{y(s)}$.  

Then if $f(t)=y(s)$ and $F(t)=x(s)$ we have
\begin{eqnarray*}
\frac{dF}{dt}=\frac{dF}{ds}\cdot\frac{ds}{dt}&=&x'(s)\div
(x'(s)/y(s))\\
&=&y(s)=f(t).\end{eqnarray*}
In particular, if our curve is the graph $(s,y(s))$ of a function, 
then we have $t=\mathlarger\int^s\dfrac 1{y(u)}\, du$.

We refer to this representation (which is essentially unique) as the
{\it canonical parametrisation} of the curve.

Note that there is no
difficulty if we confine attention to curves which lie in 
the open upper or lower half-plane and whose velocity in the direction of
the $x$-axis never vanishes.  If the curve touches or crosses
the $x$-axis or violates the second condition, then care is required and 
such a parametrisation may not
exist. (Think of  lines parallel to the $y$-axis).

We illustrate this with a very simple example:
the parabola $(s,s^2)$. This has canonical
parametrisation $(-\dfrac 1t,\dfrac 1{t^2})$.
Further examples will be computed below.

Note that the parabola, which is originally \lq \lq in one piece'',
now splits into two parts (corresponding to $t>0$ and $t<0$ respectively).
This mirrors the fact that the parabola touches the $x$-axis at its vertex.

\subsection{The duality (quadrality)}

These considerations make the following concept of duality between
curves of the above two families
natural: 
\begin{itemize}\item[A)] From spirals to parametrised curves: The
  spiral $rf(\theta)=1$ corresponds to the parametrised curve
  $\left(\mathlarger\int^tf(u)\, du,f(t)\right)$.
\item[B)] From parametrised curve to spiral: If we are given the curve
  $(x(s),y(s))$ we compute the canonical parametrisation $(F(t),f(t))$
  (i.e., $f(t)=y(s(t))$ with $t=\mathlarger\int\dfrac{x'(u)}{y(u)}\, du$) and then
  associate to it the spiral $rf(\theta)=1$.
\end{itemize}

Of course, the transition from the spiral to the parametrised curve
is immediate, while the reverse transition requires the intermediary
step of computing the canonical parametrisation.  We shall
find that this procedure sometimes leads to rather
surprising results.

We remark at this point that it is an abuse of terminology to talk
about a duality between curves since one normally does not distinguish
between congruent or even similar curves.  However, for our duality
the position of the curve with respect to the system of coordinate
axes is of crucial importance.  Thus the dual of a unit circle
with centre at the origin is very different from that of one which
passes through the origin. This mirrors the fact that the $x$-axis and
the origin respectively have a privileged role with respect to these 
two classes of curves.
Further, the dual of a curve   depends on whether we regard it as
a parametrised curve or as a \lq \lq spiral''.
Another aspect of the duality which will be of some consequence later is
the fact that although the transformation from a curve to its dual
does not arise from a transformation of space, nevertheless the transformation
{\it does} act pointwise  on the curves themselves.  Specifically, the
point $(F(t),f(t))$ on the parametrised curve is associated with 
$\dfrac{(\cos t,\sin t))}{f(t)}$ on the dual spiral.

For many purposes it will be useful to 
extend this duality 
to
a quadrality by adding the  inverse of $rf(\theta)=1$ in the unit circle
(i.e., the curve $r=f(\theta)$) and its dual, i.e., $\left(\mathlarger
\int^tg(u),g(t)\right)$ where
$g(t)=\dfrac 1{f(t)}$. Thus we have the scheme
$$\begin{array}{ccc}rf(\theta)=1&\leftrightarrow&(F(t),f(t))\\
\updownarrow&&\updownarrow\\
r=f(\theta)&\leftrightarrow&\left(\mathlarger\int\dfrac 1{f(u)}\, du,\dfrac
  1{f(t)}\right)\end{array}.$$
 
The main purpose of our note is to explain the significance of this
duality, in particular for families which are trajectories in the
above generalised sense, one with  respect to physical properties
with regard to $r^\alpha$ laws, the other with respect to $y^\alpha$ laws. 

\subsection{Examples of duality}

I. We begin with duality for spirals.  Note that we are using the term
spirals in the rather loose sense of any curve with an equation of the form
$rf(\theta)=1$.  Since the computation of the dual of a spiral requires
only an integration, it is an easy task to set up a program, say in 
Mathematica, which automatically computes the inverse and both duals
and displays plots of all four curves.  We will therefore
only mention
a few cases which are of particular interest.   
We begin with the Archimedes spiral $r=a \theta$, i.e., where 
$f(\theta)=\dfrac 1 {a \theta}$. Its inverse is
$r=\dfrac 1{a \theta}$, that is,  $f(\theta)= a \theta$, which is known
as the hyperbolic spiral.

The dual curves have parametrisations $\left(\dfrac 1 a \ln t,\dfrac 1 {a t}
\right)$ 
and $\left(\dfrac 1{a t^2},a t \right)$ respectively, 
i.e. they are the 
graph of $y = \dfrac 1 a e^{-a x}$ and the parabola $y^2 = 2 a x$.

\noindent
II.  As noted above, the canonical representation for the parabola
$(s,s^2)$ is $\left (-\frac1t,\frac1{t^2} \right)$ and so the dual spiral
is $r=\theta^2$, with inverse spiral $r\theta^2=1$ which dualises to Neil's
parabola $\left(\dfrac{t^3}3,t^2\right)$.

\noindent
III. The logarithmic spiral has the form
$r a e^{b \theta}=1$, i.e., has $f(\theta)=a e^{b \theta}$.  Its 
inverse is $\dfrac r a e^{-b \theta}=1$ and so is again a 
logarithmic
spiral.

Their duals are
the curves with parametrisations $\left(\dfrac a b e^{b t},a e^{bt}\right)$ and 
\linebreak 
$\left(-\dfrac 1{ab}e^{-bt},\dfrac 1a e^{-bt}\right)$, i.e., rays of the lines
 $y=b x$ and $y= - b x$ respectively

\noindent
IV.  The generalised parabola $y=a x^\alpha$.
In this case, the reparametrisation is
$$\left(((1-\alpha)t)^{\frac1{1-\alpha}},
((1-\alpha)t)^{\frac \alpha{1-\alpha}}\right).$$
When $\alpha<1$ the parameter $t$ ranges over the positive half-line, and
when $\alpha>1$ over the negative half-line.

This is a special case of the so-called higher-order spirals, which
are written classically as  $r^k=\dfrac \theta{2 \pi}$. (For our purposes,
$k$ can be any real number, not necessarily an integer).
Its inverse is also a higher order spiral, and the dual curves are
higher order parabolas, i.e., of the form  $y \propto x^\alpha$
for a suitable $\alpha$.

Of particular interest is the case $k=2$, which is known as the Fermat
or parabolic spiral.

Further interesting spirals are the Galilei spirals  $r=a-b \theta^2$,
the logarithmic spirals  $r = a e^{b \theta}$ and the rose curves
$r=a \sin b \theta$.  The special rose curve with $b=1$ is one of
the Cotes' spirals and is known as the epispiral.  

It is natural to
generalise the rose curves to the family  $r^d = a^d \sin (b d \theta)$ which
generalises the family of MacLaurin spirals and has some of their
propeties, albeit in a weaker form (for example, they are orbits
for force laws which are the sum of two powers ).

\subsection{The sinusoidal spirals and catenaries}

The specific properties of the sinusoidal spirals $$r^d\cos (d\theta)=1$$
suggest that their duals, i.e., the curves with
parametrisation $$(F_d(t),f_d(t))$$ where $f_d(t)=(\cos dt)^{\frac  1d}$ 
and $F_d$ is a primitive of $f_d$, will also possess interesting
properties. It turns out that this is, in fact, the case, as we shall
see shortly. We propose to call these curves {\it MacLaurin} or 
{\it sinusoidal
catenaries}
since they arise from the classical catenary using a transformation
which will  be discussed below.

The key to this phenomenon lies in the fact that the
functions $f(t)=a(\cos d(t-t_0))^{\frac 1d}$ have special
properties---namely they are solutions of differential equations of
  the form 
\begin{eqnarray*} f(t)^2+f'(t)^2&=&af^\alpha\\
f(t)+f''(t)&=&bf^\beta\\
f(t)f''(t)-f'(t)^2&=&cf^\gamma\end{eqnarray*}
as we shall see below.

The expressions $f^2+f'^2$, $f+f''$ and $ff''-f'^2$
occur frequently in computations involving the geometric and mechanical
properties of curves (we have already met the first two).  
Hence those cases where they are proportional
to a power of $f$ (and so to a power of the distance to the origin
or to the $x$-axis respectively) can be expected to have special properties.

It is not a coincidence that the above family of
functions satisfies each of these equations since the latter are
closely related, as we shall see shortly.

We note that the usual rule of thumb prepares us to anticipate
$\infty^3$ or $\infty^4$ resp. $\infty^4$ solutions for these
equations (one or two constants of integration and the
two parameters in the equation).  Thus we would expect our list to
include all of the solutions in the first case, but not in the other
two.  In fact, it is also possible to give all of the solutions
in the third case explicitly using elementary functions, and we shall do 
this in the next section.

\section{On the expressions $f^2+f'^2$, $f+f''$, $ff''-f^2$, how to solve
the equations  $f^2+f'^2 = a f^\alpha$, $f+f''=b f^\beta$, 
$ff''-f^2=cf^\gamma$
and why one might want to}

We are interested in the 
three differential
equations

\begin{itemize}\item[(A$_1$)] $f^2+f'^2=af^\alpha$;
\item[(B$_1$)] $f+f''=bf^\beta$;
\item[(C$_1$)] $ff''-f'^2=cf^\gamma$.
\end{itemize}
It will be convenient to consider the following more general situation:
\begin{itemize}\item[(A)] $f^2+f'^2=\phi(f)$
\item[(B)] $f+f''=\psi(f)$
\item[(C)] $ff''-f'^2=\rho(f)$\end{itemize}
where $\phi$, $\psi$ and $\rho$ are given smooth functions and 
the  unknown $f$ is a
function of one variable (so that our original equations correspond to
the situation where the 
functions $\phi$, $\psi$ and $\rho$ are powers of
$f$ (with a constant of proportionality).

The reason for this interest and the relevance of these equations
for geometric 
and mechanical properties of our special types of curves will
be discussed below.

The above stability property is a consequence of 
a simple computation which shows that if
$g$ is the $d$-transform of $f$, i.e.,
$g(t)=f(dt)^{1/d}$, then

\begin{eqnarray*}g^2+g'^2&=&f^{\frac 2d-2}(f^2+f'^2)\\
g+g''&=&f^{\frac 1d-2}((f^2+f'^2)+d(ff''-f'^2))\\
gg''-g'^2&=&f^{\frac 2d-2}d(ff''-f'^2).
\end{eqnarray*}
Hence if $f$ is a solution of (A), (B) or (C), then
$g$ is a solution of the corresponding equations with
the expressions $g^{2-2d}\phi(g^d)$, or 
$g^{1-2d}(\phi(g^d)g^{d}+d \rho(g^d))$ and $d g^{2-2d}\rho(g^d)$ respectively 
as the 
right-hand sides.

\subsection{The relationship between the equations}

These equations and their solutions are closely related, as we shall
now see. 

\noindent
1) If $f$ is a solution of (A), i.e., if $f^2+f'^2=\phi(f)$, then,
differentiating and simplifying, we get
$$2ff'+2f'f''=\phi'(f)f'$$
and so $f+f''=\frac 12\phi'(f)$, i.e., $f$ is a solution of (B) with
$\psi(f)=\frac 12\phi'(f)$. 
\noindent
Also $1+\dfrac{f'^2}{f^2}=\dfrac{\phi(f)}{f^2}$, and so, again by
differentiating and simplifying, we get

$$(ff''-f'^2)=\frac 12 f \phi'(f)- \phi(f),$$ 
i.e., $f$ is a
solution of (C) with $\rho(f)=\frac 12f \phi'(f)-\phi(f))$.

\noindent
Now assume this $f$ is a solution of (B), i.e., $f+f''=\psi(f)$. Then by
the above $f^2+f'^{2}=2\phi(f)$, where $\phi$ is a primitive of
$\psi$. Hence
$$ff''-f'^2=\frac 12f\psi(f)-\phi(f),$$
an equation of the form (C).
\noindent
Finally, if $f$ is a solution of (C) then we can solve the 
(linear) differential
equation
$$\frac 12 f \phi'(f)- \phi(f)=\rho(f)$$
for $\phi$, given $\rho$, and so obtain an equation of type (A) for
$f$. 

\subsection{Some solutions}

In fact, using standard elementary techniques for solving O.D.E.s by
quadrature, we can express the solutions implicitly as follows:

\begin{tabular}{rll}(A)&$f^2+f'^2=\phi(f)$&$f(t)=F^{-1}(t+c)$, where
  $F$ is a primitive of $(\phi(u)-u^2)^{-\frac 12}$\\
(A$_1$)&$f^2+f'^2=af^\alpha$&special case of (A)\\
(A$_2$)&$f^2+f'^2=af^\alpha+d$&special case of (A)\\
(B)&$f+f''^=\psi(f)$&special case of (A) with $\phi=2\mathlarger\int\psi$\\
(B$_1$)&$f+f''=bf^\beta$&special case of (B)\\
(C)&$ff''-f'^2=\rho(f)$&$f(t)=\exp(G^{-1}(t+c))$, i.e., $\ln f(t)=G^{-1}(t+c)$\\
&& where $G$ is a primitive of
$(2\mathlarger\int\rho(e^u)e^{-2u})^{-\frac 12}$\\
(C$_1$)&$ff''-f'^2=cf^\gamma$&special case of (C). 
\end{tabular}

\noindent
We repeat for the record that
if $f(t)=p(\cos d(t-t_0))^{\frac 1d}$ then \\
\begin{eqnarray*}f^2+f'^2&=&p^{2}f^{-2d+2}\\
f+f''&=&-p^{2}(d-1)f^{-2d+1}\\
ff''-f'^2&=&-p^{2}df^{-2d+2}.
\end{eqnarray*}

\noindent
Thus this $f$ solves (A$_1$) with $a=p^{2}$, $\alpha=-2d+2$\\
(B$_1$) with $b=-p^{2}$, $\beta=-2d+1$\\
and (C$_1$) with $c=-p^{2}$, $\gamma=-2d+2$.

\noindent
It is interesting that the hyperbolic trigonometric functions have similar
but not identical properties, namely if $g(t)=p(\cosh d(t-t_0))^{\frac 1d}$, 
then\\
\begin{eqnarray*}g^2-g'^2&=&p^{2}g^{-2d+2}\\
g-g''&=&-p^{2}(d-1)g^{-2d+1}\\
gg''-g'^2&=&p^{2}dg^{-2d+2}.
\end{eqnarray*}

\noindent
We will mention a possible application of these facts below.

The noteworthy form of the above solutions shows that they all arise
from the simple function $\cos t$ by applying the $d$-transformation
$f(t)\leadsto f(dt)^{\frac 1d}$ which, we recall, corresponds to the 
transformation $z \mapsto z^d$  of the complex plane. This fact
and, in particular, the fact that the latter  is conformal,
is relevant for some of the remarkable properties of the families
(see, e.g. [Sc2] where this is used to explain many properties
of the MacLaurin spirals).

The reason for the above stability property lies in
a simple computation which shows that if\\
$g(t)=f(dt)^{1/d}$, then\\
\begin{eqnarray*}g^2+g'^2&=&f^{\frac 2d-2}(f^2+f'^2)\\
g+g''&=&f^{\frac 1d-2}((f^2+f'^2)+d(ff''-f'^2))\\
gg''-g'^2&=&f^{\frac 2d-2}d(ff''-f'^2).
\end{eqnarray*}

Hence if $f$ is a solution of (A), (B) or (C), then $g$ is a
solution of $g^2+g'^2=g^{2-2d}\phi(g^d)$,
or $g+g''=g^{1-2d}(\phi(g^d)+d\rho(g^d))$,
or $gg''-g'^2=dg^{2-2d}\rho(g^d)$.

In order to motivate these considerations we shall now show that the
above three differential expressions are ubiquitous in calculations which
are related to mechanical or geometric properties of curves.

\subsection{Where the expressions arise}
As mentioned above we have two central motivations for these considerations.
Firstly, the expressions of our 
title occur in many formulae which arise
in analytically describing geometrical or mechanical properties of 
curves described as above.  Secondly, and as a consequence, for 
curves which correspond to functions
with the property that they are solutions of the above equations, 
then these quantities will take on a particularly simple
form and so the curves will have remarkable geometric and mechanical
properties.  We believe that this is the natural explanation of the
special properties of the MacLaurin spirals and catenaries.

We bring a sample of the kind of occurence that we have in  mind:

\noindent
For the spiral  $r f(\theta) =1$, then

\begin{enumerate}
\item
the curvature is $\dfrac{f^3(f+f'')}{(f^2+f'^2)^{\frac 32}}$;

\item
the affine curvature is $f^3(f+f'')$ (see below);

\item
if a planet moves around the curve in such a manner that it obeys
Kepler's second law, then the absolute value of the acceleration
vector is $f^2(f+f'')$;

\item
the curvature of the hodograph is $(f+f'')^{-1}$;

\item
the infinitesimal length is $ds^2 = \dfrac{1}{f^4}(f^2+f'^2)d\theta^2$.
\end{enumerate}

\noindent
For the curve with parametrisation $(F(t),f(t))$,

\begin{enumerate}
\item
the curvature is $\dfrac{ff''-f'^2}{(f^2+f'^2)^{3/2}}$;

\item
if it moves under a force parallel to the $y$-axis which depends only
on the distance to the $x$-axis, then the absolute value of its
acceleration is $\dfrac{ff''-f'^2}{f^3}$;

\item
The infinitesimal length is given by the formula $ds^2=(f^2+f'^2)dt^2$.
\end{enumerate}
A further situation where the expression $f^2+f'^2$ occurs will now 
be discussed briefly.

\subsection{The calculus of variations}

One of the most important aspects of the sinusoidal spirals and catenaries
is that they provide a plethora of solutions to natural problems 
of the variational calculus.
Elementary treatises on this subject typically use the
following model applications, which are mathematically interesting,
are of great historical interest and can be computed explicitly:
the catenary (the curve which minimises potential energy
and whose surface of revolution
has a minimal-area property), the cycloid (brachistochrone), Dido
circles (curves of a given length with endpoints on the $x$-axis
which enclose maximal area).  We shall now show that these are special 
cases of a general
phenomenon and that this explains many of the remarkable properties
of the MacLaurin spirals and catenaries.

The common structure of these problems is that they maximise or minimise
functionals of the form

$$\mathlarger\int y^\delta \,ds \quad \text{or} \mathlarger\int r^\gamma \, ds.$$

\noindent
We consider the second case:
If we use our basic parametrisation

$$
\frac{(\cos(t),\sin(t))}{f(t)} \quad $$
then it reduces to the variation problem of
minimising or maximising a functional of the form

$$\mathlarger\int F_\beta(x,f(x),f'(x))\,dx$$
where  $F_\beta (x,u,v) = u^\beta(u^2+v^2)^{\frac 12}$
for a suitable $\beta$ ($\beta=\delta$ resp. $\beta=-\gamma-2$).
We now use the simple fact that for each  $\beta$,
$f$ is a solution
of the corresponding Euler-Lagrange equation if and only if  
$f^2+f'^2 = a f^\alpha$ for some $a$
where $\beta+2=\dfrac\alpha 2$. 
This is seen by using the standard fact that in the case, as here, where
the corresponding integrand is independent of the first variable, the
Euler equation has the form:  $vf_3-f$ is constant.  This expression is
$$\frac{u^{\beta+2}}{a^{\frac 12}u^{\frac \alpha 2}} $$ 
when $f^2+f'^2=a f^\alpha$, and so
is constant when the above relationship holds.

This explains many of the remarkable
mechanical properties of the MacLaurin spirals.  

For example, in [Br]  it is shown that the $\infty^2$ 
family of parabolae of the form
$$  \alpha(y-\alpha)=(x-d)^2$$
are the paths of light rays in a medium with refraction index 
$\propto y^{\frac 12}$.
This is a special case of the above result since the canonical parametrisation
of the above curve is $$\left  (2 \alpha \tan \frac t 2+d,
\alpha \left (\cos \frac t 2\right )^{-2}\right )$$
and for $f(t)=\left(\cos \frac t 2\right )^{1/2}$, we have
$ f^2+f'^2=f^3$.

We now show that the MacLaurin catenaries are
also solutions of suitable variational problems.  
Interestingly, although the concrete variational problems which are usually
featured as text book examples are usually formulated for the case of
a parallel force, many more concrete examples are known for the
case of a central one.
However, using the methods 
expounded here, we can extend the list of such examples for parallel forces
considerably.

We consider the problem of maximising

$$\mathlarger\int y^\delta \,ds$$

i.e. the variational problem with kernel

$$J(y,y')=y^r(1+y'^2)^{-\frac 12} .$$

The corresponding Euler equation reduces to

$$y^r(1+y'^2)^{-\frac 12} =c $$

where $c$ is an arbitrary constant.

If we now use the parametrisation
$(F(t),f(t))$,
this reduces to $f^{r+1}(f^2+f'^2)^{-\frac 12}=c$.
 
Hence if $f^2+f'^2$ is proportional  
to $f^\alpha $, then the above parametrisation
provides a solution whenever $r+1=\frac \alpha 2 $.  Hence we obtain
a solution  with $f(t)=p (\cos d t)^{\frac 1d}$ when $2-2d=\alpha $.
\subsection{Two more subtle examples}

As we have seen above, Dido circles and cycloids are examples of 
MacLaurin's catenaries. But why are catenaries, i.e., the graphs of the
functions  $a \cosh \left(\frac xa \right)$ also members of this family,
as the name implies?  This is because they have canonical parametrisations
$(F_{-1}(t),f_{-1}(t))$ as the reader can verify.

A further fascinating class of curves are formed by the two parameter
family of parabolae of the
special form $y= a+\frac{(x-d)^2}{4a}$, which are the paths taken by light rays
in suitable media.  Once again, these are MacLaurin catenaries, explicitly
$(F_{-2}(t),f_{-2}(t))$.  This explains the following remarkable fact
from [2]: The above family of parabolae are the paths followed by light
rays in a medium where the index of refraction $n(x,y)$ is proportional
to $y^{-\frac12}$.  Our method allows us to solve the corresponding problem
for $n(x,y) \propto y^\alpha$ for any $\alpha$.

\subsection{Miscellanea}

In this section we bring some miscellaneous remarks on our main
topic.

\subsubsection{The explicit form of the parametrisation
  $(F_d(t),f_d(t))$ }

The reader will have noticed that we never required the explicit form of the
primitive of $f_d$ and so the explicit parametrisation of the MacLaurin
catenaries.  For the record we include it here: \\
$x(t)=-\frac 1{1+d}(\cos(dt)^{1+\frac 1d}\mbox{\rm cosec}(dt)F_1^2(\frac
12(1+\frac 1d).\frac 12,\frac 12(3+\frac 1d),(\cos(dt)^2)$\\
$y(t)=(\cos^d(dt))^{\frac 1d}$\\
(we used Mathematica to obtain this expression---$F^2_1$ is the hypergeometric 
function.)

\subsubsection{The connection with contact geometry} 
The representation of a curve in the form $(F(t),f(t))$, 
which might seem rather artifical and {\it ad hoc},
is
natural within the framework of contact geometry as was pointed out to
me by Valentin Lychagin. Namely, if we consider the simplest contact manifold
${\bf R}^3=\{(x,y,t):x,y,t\in{\bf R}\}$ with the natural
structure given by the differential form $\omega=dx-y\, dt$, then
the Legendrian submanifolds are the curves of the form $((F(t),f(t),t)$
i.e. our special parametrisation displays the curve as the projection
of such a manifold onto the $(x,y)$ coordinates.

\subsubsection{Curves with prescribed curvature}

Many problems in applied mathematics can be subsumed under the general
one: given a function $f(x,y)$ on the plane or a subset thereof, we
are required to find a parametrised curve such that at each point
$(x,y)$ on the curve the curvature $\kappa$ there is given by $\kappa=f(x,y)$. 
In
other words we are looking for solutions of the equation
$$\frac{\dot x(t)\ddot y(t)-\ddot x(t)\dot y(t)}{(\dot x(t)^2+
\dot y(t^2))^{\frac 32}}=f(x(t),y(t)).$$  
It follows from above that the sinusoidal spirals are solutions
of this problem for the case $\kappa \propto r^\alpha$ for corresponding
$\alpha$.  This fact, which was known to the classical differential geometers
(cf. [Go], [Lo],
has been rediscovered (for special cases)  
several times (cf. e.g. [Si]).
Thus MacLaurin had shown that the radius of curvature of the  spiral
is given by the formula

$$R=\frac{a^n}{(n+1)r^{n-1}}. $$

Special cases of these reults can be found in the modern literature.
All these results follow easily from the theory developed here.
Analogously, the Maclaurin catenaries are solutions of related problems
for $\dfrac 1{y^\alpha}$ laws.

All these results follow easily from the theory developed here.
Analogously, the Maclaurin catenaries are solutions of related problems
for $\dfrac 1{y^\alpha}$ laws, as we shall show below.

\subsubsection{The rectification problem}

If $(x(t),y(t)=\dfrac{(\cos t\sin t)}{f(t)}$ is the canonical
parametrisation of the curve $rf(\theta)=1$ then $ds^2=\dfrac
1{f^4}(f^2+f'^2)\, dt^2$.
If $(x(t),y(t))=(F(t),f(t))$ then
$ds^2=(f^2+f'^2)\, dt^2.$
For $\left(\mathlarger\int^t\dfrac 1{f(u)}\, du,\dfrac 1{f(t)}\right)$ we have
$ds^2=\frac 1{f^4}(f^2+f^{12})\, dt^2.$
Hence the expression for the infinitesimal length of the spiral
$rf(\theta)=1$ and for its \lq \lq diagonal dual'' 
$(G(t),g(t))$, where $g=\dfrac 1f$ and $G$ is a primitive of $g$,
coincide.  This implies that the length between two points
on the spiral and the corresponding ones on the diagonal dual
are the same.  This is the background for many rectification 
results in the classical
theory.
For example, if $r=a\theta$, $f(\theta)=\dfrac 1{a\theta}$ and so the
dual curve is $\left(\dfrac 1a\ln t,\dfrac 1{at}\right)$, i.e., the logarithmic
spiral $y=\frac 1ae^{-ax}$. The polar reciprocal is $ra\theta=1$ with dual
$\left(\dfrac{at^2}2,at\right)$, i.e., the parabola $y^2=2ax$.
In the words of [Lo] this means that the rectification problem
for the Archimedean spiral is equivalent to the analogous problem
for the parabola.
The rectification problem played an important role in the history of
geometry, for which see the same reference. Since this question seems rather
innocuous from a modern point of view, it is perhaps of interest to
recall the historical perspective as described there:  
\begin{quote}Thus we see that Archimedes had solved two of the three fundamental
questions for his spirals, namely the determination of its tangent
and of the area swept out by a radial vector moving along the curve,
but did not touch on the rectification problem.  This important
task 
could not be solved by his contemporaries or immediate successors,
and would only be successfully attacked nearly two thousand years
later, independently by Cavalieri, St. Vincentius, Roberval, Pascal
and Fermat'' (our translation). 
\end{quote}
The method used here immediately givey a plethora of such results.
We mention only one such, due to Fermat:  the rectification
problems for the spiral  $r^k=\dfrac{a^k\theta}{2 \pi}$
and the generalised parabola  $z^{k+1}=(k+1)p^kx$,
when $a^k=2 k \pi p^k$ ([10],p. 435), are equivalent.

\subsubsection{The Weingarten mapping of surfaces of revolution}

We consider the surface of revolution
$$\phi(u,v) = 
(f(v)\cos u ,F(v),f(v) \sin u)$$
obtained by rotating the curve $(0,F(v),f(u))$ around the $x$-axis.
Then if we compute the coefficients of the two fundamental forms, 
we find that 

$$E= f^2, \qquad F=0, \qquad G= f^2+f'^2$$
and $$L = f^2/(f^2+f'^2)^{1/2},\qquad M = 0, \qquad 
N=(f'^2-ff'')/(f^2+f'^2)^{1/2}.$$ 

Once again, we see the prevalence of the three expressions
$$f^2+f'^2, \quad f+f''\quad \text {and } ff''-f'^2.$$

As a consequence, if we rotate the MacLaurin catenary $(F_d(t),f_d(t))$
around the $x$-axis, then all of the quantities involved, i.e., $E$, $G$,
$L$ and $N$ (and, as a consequence, the principal curvature and Gau\ss ian
curvature), are proportional to powers of the distance from the $x$-axis.
($F$ and $M$ vanish, of course).

Without going into details we note that the Christoffel symbols
for these surfaces (and hence the geodetic equations) also take on a very
simple form.

If instead of using a classical
Euclidean rotation we employ a hyperbolic one, i.e. the linear mapping
with matrix

$$\left[\begin{array}{rr} \cosh\theta&\sinh\theta\\
\sinh\theta&\cosh\theta\end{array}\right],$$
we obtain corresponding results for the Minkowski
pseudo-metric (i.e., the  one corresponding to the 
quadratic form $x^2+y^2-z^2$),
which suggests that such surfaces could
be of interest in the theory of relativity (this follows from the relationships
satisfied by the hyperbolic trigonometric functions displayed above).

\subsubsection{Remark}
The characterisation of the functions $p(\cos(d(t-t_0))^{1/d}$ 
as solutions of an equation
of the form $ f^2+f'^2=a f^\alpha $ is implicit in MacLaurin's treatment.
He defines the spirals as curves which satisfy the condition
$$\tan V=r \frac{d\theta}{dr}, $$
where $V$ is the angle between the vector from the origin to the point 
$ (r,\theta)$
on the curve and the tangent there.  This condition is equivalent to the
differential equation
$$\frac{d\theta}{dr}=\frac{r^{n-1}}{\sqrt{a^{2n}-r^{2 n}}} $$
(cf. Gomes Teixeira [Go], p. 259,  but note that there is a misprint in the 
equation there). If we put $f(\theta)=\dfrac 1 {r(\theta)} $, then the
above equation reduces to $$ f^2+f'^2=a^{2 n}f^{2n-2}.$$

\section{The trajectories for parallel power laws}

We now turn to a special case of the following problem: given a force field
(in our examples on a subset of the plane), determine the family of 
all trajectories of particles moving under this field.
An analogous problem is: given a function on a subset of the plane, determine
the two-parameter family of curves which are such that at each point 
on the curve, the curvature corresponds to the value of this function.
(Such problems arise, e.g., in theory of elasticae---cf. [Si]).
We will give explicit formulae for the 
genuinely planar trajectories
of particles which move under a force law of the form $K \propto y^{\alpha}$.
In a second section we investigate free fall
along a line parallel
to the $y$-axis for such laws. 
In the final section we show that analogous methods can be used
to solve the curvature problem, again for the case where the curvature
is proportional to a power of the distance from the $x$-axis.
We are using the term \lq \lq explicit'' in the following precise sense.
In the first case, we show that the trajectories can be parametrised
by functions which are called elementary in  normal usage (we include the
hypergeometric functions in this category).
In the two second cases, we have to allow the use of the inverses of such
functions.

More precisely, we shall show that the trajectories for a $y^\alpha$-law 
are the curves with 
parametrisations $(F_d(t),f_d(t))$,
where $f_d$ is a function of one of   four particular forms 
for suitable parameters $a$, $b$ and $c$,  $F_d$
is a primitive of $f_d$  and $d$ is $\alpha+1$,
where $\alpha$ is the index of the power law.

The four forms of $f_d$ are
\begin{eqnarray*}
f_d(t)=\left(\frac{1}{a d^2 t^2}\right)^{\frac{1}{d}};
\end{eqnarray*}
for fixed $d$, this generates
a two-parameter or $\infty^2$ family of curves (the parameters are
$a$ and  an integration constant).  The parameter  $t$ ranges over
the punctured line, i.e.  $\bf R \setminus \{ 0\}$.
($f_d$ is, of course, proportional to  $t^{-\frac 2d}$, but we 
leave it   in the above
form for reasons which will soon become obvious).

\begin{eqnarray*} 
f_d(t)=\left(\frac{a (b-c)^2 e^{a (b+c) d t}}{\left(e^{a b d t}+e^{a c d
   t}\right)^2}\right)^{\frac{1}{d}},
\end{eqnarray*}
a three-parameter family.  Here $t$ ranges over the whole real line.

\begin{eqnarray*}
f_d(t)=\left(\frac{a (b-c)^2 e^{a (b+c) d t}}{\left(e^{a b t}-e^{a c d
   t}\right)^2}\right)^{\frac{1}{d}},
\end{eqnarray*}
a three-parameter family.  Once again, $t$ ranges over the punctured line.

\begin{eqnarray*}
f_d(t)=\left(a c^2 \sec ^2\left(a c d t\right)^2\right)^{\frac{1}{d}},
\end{eqnarray*}
a three-parameter family  ($a$, $c$ and an integration constant).
$t$ ranges over the set of values for which   $a c d t$ lies in the 
open interval $]-\frac \pi 2,\frac \pi 2 [$.

The reader will have noticed that there is a problem with a $\dfrac 1 y$
law, i.e. where $d=0$ and we discuss this case later.

These are parametrisations (in the differential geometric sense)
of the curves traced out by a particle moving under the corresponding
force law---they do not describe its actual
motion, i.e.,  $t$ is not time.  The parametrised motion
is  $(u,f\circ F^{-1}(u))$, and cannot, in general, be
written out explicitly using elementary functions but also requires the inverse
of such a function.

We think that it is of some interest to document the fact that we
can write down explicitly {\it all} the trajectories, using simple elementary
functions (and an integration), in the case of a parallel power law.
We remark that these integrations can be carried out by Mathematica
and the results can also be expressed in terms of elementary functions 
(if one admits this 
status to the hypergeometric functions).  We have added the explicit forms
at the end of the section.   
In contrast to the case of parallel laws, there are, for central
power laws, just three cases in which all of the trajectories can
be described explicitly using elementary functions: the Kepler case
$K \propto r^{-2}$ mentioned above, Hooke's law $K \propto r$ (where
the orbits are conic sections with centre at the origin) and the so-called
Cotes' spirals ($K \propto r^{-3}$).  In general, the 
trajectories for a given force law form a three-parameter or $\infty^3$
family.  For the general central power law MacLaurin
produced an explicit $\infty^2$ family of trajectories (later baptised
as the MacLaurin spirals---cf. [Co1]),  but the remaining ones can only be 
described
indirectly in the general case as far as we know 
(using functions which can be determined
implicitly after a quadrature).  Thus in the Kepler case, MacLaurin's
family only picks up the parabolic orbits.

The special case of rectilinear motion turns out, perhaps surprisingly,
to be more intricate, and it is interesting to note that Newton, in his
Principia, devoted a whole section to this case, which he regarded as
a limiting case of the planar one (for a central force).  In this case, it is, of course, not the
geometrical form of the motion which is of interest, but  its direct
description, i.e., formulae for the position as a function of time.
Here the results are less
satisfactory in the sense that we have to use not just elementary
functions but also the inverse of such a function.

For a brief introduction to the curvature problem and its relationship
to the Euler elasticae, see Singer [Si].
A similar proviso applies to our solutions here.
\subsection{Planar motion under a parallel law}
We will now briefly discuss the method which leads to the above formulae.
Once one has them, then it is a routine application of the chain
rule to show that the curves they describe actually have the claimed
property, i.e., that they are trajectories for suitable laws.  The only 
interesting fact is to show how these formulae arise, and this we shall
now do. Since we are considering
$y^\alpha$ laws, then it is natural to confine attention to curves in the
upper half-plane.  In the Galilean
situation, we must consider four cases, depending on the position 
of the parabola with respect to the $x$-axis, exemplified by the graphs
of the following functions:

\begin{eqnarray*}
y=x^2
\end{eqnarray*} 

\noindent
(a full parabola touching the $x$-axis);

\begin{eqnarray*}y=x(1-x)
\end{eqnarray*} 

\noindent
with $x$ between $0$ and $1$ (an arc of a parabola with
vertex pointing upward);

\begin{eqnarray*} y=x(x-1)
\end{eqnarray*}  
with two unbounded branches corresponding to $x<0$ and $x>1$;

\begin{eqnarray*}y=x^2+1
\end{eqnarray*} (a parabola which is disjoint from the $x$-axis).

Of course, only 6) is physical, i.e., corresponds to an attractive 
force towards the $x$-axis.  The other three describe trajectories
under a repulsive force (anti-gravity).

We remark firstly that in this section we are excluding the cases of
free fall parallel to the $y$-axis, which implies that we can parametrise the
trajectories in the form $(F(t),f(t))$, where $F$ is a primitive of $f$,
as shown above.
Our proof is a combination of the following simple observations:

\begin{itemize}

\item[A)]  If we parametrise a curve in the special form $(F(t),f(t))$ as above
(i.e. where $F$ is a primitive of $f$) and a particle moves along
this curve under a parallel force law, then the acceleration at the 
point $(F(t),f(t))$ is $\left(0,\dfrac {f(t)f''(t)-f'(t)^2}{f^3(t)}\right)$.

\noindent
This is a simple consequence of the chain rule.
\end{itemize}
\noindent
Hence
\begin{itemize}
\item[B)]  We have a $y^\alpha$ law if and only if $f$ satisfies an equation of
the form  $f(t)f''(t)-f'(t)^2 = b f(t)^\beta$ for constants  $b$ and
$\beta$.  The relationship between this $\beta$  
and the exponent $\alpha$ of the power law is very 
simple---$\beta = \alpha + 3$.  Of course, the force is 
attractive or repulsive (with reference to the $x$-axis)
according  to whether  $b$ is negative or
positive.

\item[C)]  If  $f$  satisfies an equation as in B), then so does
its $d$-transform
$f_d$ (with distinct parameters $b$ and $\beta$).
This follows from the equation:
$$f_d f_d''-f_d'^2=d f^{-2+\frac2d} (ff''-f'^2)$$
which we gave above.
\end{itemize}
\noindent  The crucial point is now the following rather unorthodox 
parametrisations of the parabola:
\begin{itemize}
\item[D)]  The general parabola (more precisely,  the part 
or parts above the $x$-axis)
has canonical parametrisation $(F(t),f(t))$
where  $f$  has one of
following four forms:
\end{itemize}

\begin{eqnarray*}
f(t)=\dfrac{1}{a t^2};
\end{eqnarray*}
$(F(t),f(t))$ is then a parametrisation of the parabola  $y=a(s-b)^2$ 
where we choose
$F$ so that $F(0)=b$. (Again, $t$ ranges over the punctured reals and the
vertex is \lq \lq lost'');

\noindent or

\begin{eqnarray*}
f(t)=\frac{a (b-c)^2 e^{a (b+c) t}}{\left(e^{a b t}+e^{a c t}\right)^2};
\end{eqnarray*}
$(F(t),f(t))$ is then  a parametrisation of the parabola $y=a(x-b)(c-x)$, 
ormore precisely
of the arc in the upper half plane. $t$ ranges over the real line;

\noindent
or 

\begin{eqnarray*}
f(t)=\frac{a (b-c)^2 e^{a (b+c) t}}{\left(e^{a b t}-e^{a c t}\right)^2};
\end{eqnarray*}

\noindent
$(F(t),f(t))$  then parametrises $y=a(x-b)(x-c)$, or 
more precisely the two unbounded branches
in the upper half plane.  They correspond to $t<0$ and 
$t>0$ respectively;

\noindent
or

\begin{eqnarray*}
f(t)=a c^2 \sec ^2(a t)
\end{eqnarray*}

\noindent
where $(F(t),f(t))$ parametrises $y=a((x-b)^2+c^2)$.  $t$ ranges over the open
interval $]-\frac\pi{2a},\frac\pi{2a}[$.

\noindent
(The four  different forms correspond to the four possible positions
of the parabola with regard to the  $x$-axis, as described above).

\noindent
In these parametrisations, $a$ is assumed to  be positive and $b<c$.

\begin{itemize}
\item[E)] in each of the cases in D) the function  $f$ satisfies the condition  
$$f(t)f''(t)-f'(t)^2 =
2a f(t)^3.$$
\end{itemize}
We have thus verified Galilei's result that the parabolas are
orbits under a constant parallel force law, albeit in a rather roundabout 
manner.  
However, the payoff from this approach now follows.  
\begin{itemize}
\item[F)] If we combine this
fact and the formula in C) then we see that the functions quoted at the
beginning
of the article (which arise from the above parametrisations of the parabola
via the process considered there) satisfy the 
condition that $ff''-f'^2$ is proportional to a power of
$f$, in fact to $f^{2+d}$,
and so the curves mentioned in the first paragraph are trajectories
for $K \propto
y^{d-1}$.  Of course, as mentioned above, 
one can compute this directly; but the treatment
described here shows how to find the appropriate form for the functions $f$.
\end{itemize}

We now obtain the results above as follows.  We suppose that we have
a trajectory for a $y^\alpha$ law parametrised in the form
$(G(t),g(t))$.  Then for any $\alpha$ we can find
a suitable  $d$  so that $g(t)=f_d(t)=f(d t)^{1/d}$, where  $f$
corresponds to the Galilean case, i.e., a constant force.  We then
use the above formula for the possible forms for $f$ and so find
$g$ as above.

It thus only remains to consider the singular case, that of a $\dfrac 1y$ law.
In this case, the differential equation for $f$ (i.e. $f f''-f'^2=cf^2$)
can be solved directly,
and one sees that the trajectories are  $(F(t),f(t))$,
where  $f$ is a function of the form  $\exp p(t)$ with $p$  a 
quadratic function.
  
For the sake of completeness, we add the explicit forms of the functions
$F_d$, i.e., the primitives of the corresponding $f_d$  
for cases 2), 3) and 4) at the head of the article (the case 1) is trivial):
\begin{eqnarray*}
F_d(t)&=&\left(\frac{e^{d t}}{\left(1+e^{d t}\right)^2}\right)^{\frac{1}{d}}
   \left(1+e^{d t}\right)^{2/d} \,
   _2F_1\left(\frac{1}{d},\frac{2}{d};1+\frac{1}{d};-e^{d t}\right)\\
F_d(t)&=&\frac1{a(b-c)}\left(\left(1-e^{a (b-c) d t}\right)^{2/d} \left(\frac{a (b-c)^2 e^{a
   (b+c) d t}}{\left(e^{a b d t}-e^{a c d
   t}\right)^2}\right)^{\frac{1}{d}}\right. \\
&&\left. _2F_1\left(\frac{1}{d},\frac{2}{d};1+\frac{1}{d};e^{a (b-c) d
   t}\right)\right)\\
F_d(t)&=&\frac1{2 a d}\left(\cos ^2(a d t)^{\frac{1}{d}-\frac{1}{2}} \,
   _2F_1\left(\frac{1}{2},\frac{1}{2}+\frac{1}{d};\frac{3}{2};\sin ^2(a d
   t)\right)\right.\\
&&\left. \left(a c^2 \sec ^2(a d t)\right)^{\frac{1}{d}} \sin (2 a d
   t)\right).
\end{eqnarray*}

\noindent
These expressions were computed with Mathematica
(in the first equation we have assumed that $a=1$, $b=0$ and $c=1$ to make the
formula more tractable).

In the singular case (i.e. a $\dfrac 1y$ law), the formula
for $F(t)$ requires the use of the error function from statistics.
Since the explicit form depends on the nature of the quadratic
function, i.e., on whether it is definite or indefinite, we omit the
details.

As final remarks, we note firstly that despite the complex form of these 
formulae,
they do, of course, specialise to simple curves for certain choices of 
the parameters.  For example a Dido circle (that is, a circle with
centre on the $x$-axis) is a trajectory for a $\dfrac 1{y^3}$
law (interestingly, this case was treated by Newton
in his Principia).
Secondly, using the theory developed here, it is easy to derive a criterion
for a given curve with parametrisation $(x(s),y(s))$
to be the trajectory of a power law.  It is that $\dfrac {A(s)}{B(s)}$
be constant, where
\begin{eqnarray*}
A(s)&=&\left(3 y'(s)y''(s)+y(s)y'''(s)\right)
x'(s)^2-\left(3 x''(s)y'(s)^2+y(s)x'''(s)y'(s)\right.\\
&&\left.+3 y(s)x''(s)y''(s)\right)x'(s)+y(s)y'(s)x''(s)^2\\
B(s)&=&{x'(s)y'(s)\left(x'(s)y''(s)-x''(s)y'(s)  \right)}.
\end{eqnarray*}

The index of the power law is then this constant minus three.

Using this criterion one can check that not only are Dido circles trajectories
for the $y^{-3}$ power law, but also that, as mentioned above, they 
are the only circles which
are trajectories for {\it any}
power law.  Ellipses and hyperbolae (not necessarily rectangular) 
with the $x$-axis as axis are also
trajectories for $y^{-3}$ laws.

\subsection{Rectilinear motion}
We now consider the case of rectilinear motion. This was   
treated by Newton in Section VIII of his Principia.  Interestingly,
he did not deal with it directly but considered it as a limiting case
of planar motion under a central force.  We shall do the same here,
with the difference that we shall employ a parallel force.

Firstly we note that this is equivalent to solving a differential equation
of the form  $$y''=F(y)$$
for the specal case where  $F$ is a power function.  Now there is a standard
method for solving equations of the above type by quadrature: we introduce
the variable  $w$ where $y'=w(y)$. Then $y''=w\dfrac{dw}{dy}$,  so the 
equation
reduces to $\dfrac{d}{dy} w^2=2F(y)$
and so we have
$w^2=2 F_1(u)\,du$, i.e.,
$w=\sqrt{2 F_1(u)\,du}$ where $F_1$ is a primitive of $F$.

Thus we have reduced to the equation
$y'=\sqrt{2  F_1(u)\,du}$,
which we can solve by quadrature.
(This is essentially the modern version of Newton's treatment as exponded
by Chandrasekhar [Ch]).
Using the methods of the first section, 
we can give a  more direct approach which does
not require quadratures.

If we suppose that  $f$ is a function which satisfies the condition
$$ f f''-f'^2=af^\alpha$$ and we set $y(t) =f\circ F^{-1}(t)$,
(with $F$ again a primitive of $f$), then the computations above
show that $$\frac{d^2 y}{dt^2}=a y^{-3} $$
and the same argument as we used above shows that we obtain all of the
solutions in this manner.  Once again, we can choose $f$ as one of the
special cases above and thus obtain analytic expressions for rectilinear
motion under a parallel power law.

The case of a $\dfrac 1y$ law is again an exception.

\end{document}